\pdfoutput=1 
\documentclass{JINST}

\title{CosmoDM and its Application to Pan-STARRS data}

\author{S. Desai$^{a,b}$\thanks{Corresponding author.}, J. J. Mohr$^{a,b,c}$, R. Henderson$^a$, M. K\"{u}mmel$^a$, K. Paech$^a$,   and M. Wetzstein$^a$ \\
\llap{$^a$}Department of Physics, Ludwig-Maximilians University, Scheinerstr.\ 1, 81679 Munich\\
\llap{$^b$}Excellence Cluster Universe, Boltzmannstrasse 2, 85748 Garching\\
\llap{$^c$}Max Planck Institute for Extraterrestrial Physics, Giessenbachstrasse 2, 85748 Garching\\
E-mail: \email{shantanu@usm.lmu.de}}

\abstract{The Cosmology Data Management system (CosmoDM) is an automated and flexible  data management system for the processing and calibration of data from optical photometric surveys.  It is designed to run on supercomputers and to minimize disk I/O to enable scaling to very high throughput during periods of reprocessing.  It serves as an early prototype for one element of the ground-based processing required by the Euclid mission and will also be employed in the preparation of ground based data needed in the {\it eROSITA} X-ray all sky survey mission.  CosmoDM consists of two main pipelines. The first is the single-epoch or detrending pipeline, which is used to carry out the photometric and astrometric  calibration of raw exposures.  The second is  the co-addition pipeline, which combines the data from individual exposures into deeper coadd images and science ready catalogs.  A novel feature of CosmoDM  is that it uses a modified stack of Astromatic software which can read  and write tile compressed images.  Since 2011, CosmoDM has been used to  process data from the DECam, the CFHT MegaCam  and the Pan-STARRS cameras. In this paper we shall describe how processed Pan-STARRS data from  CosmoDM has been used  to  optically confirm and  measure  photometric redshifts of Planck-based  Sunyaev-Zeldovich effect selected cluster candidates.}


\begin{document}

\section{Introduction}\label{sec:xxx}

To address questions in cosmology at the interface of fundamental physics such as the nature of dark energy, dark matter, and the origin of inflation, a number of astronomical photometric surveys have begun mapping out large portions of the sky at multiple wavebands. For these surveys, brand-new state of the art Mega-pixel cameras have been designed that take advantage of technological breakthroughs in deep-depletion CCD technology and fast readout electronics.  These surveys, which have been classified as Stage-III dark energy experiments by the Dark Energy Task Force,  are expected to probe the dark energy equation of state parameter using a combination of techniques such as galaxy cluster counts, weak lensing, BAO, and Type 1a SN.  The expected accuracy from the Dark Energy Survey (DES), which is one such Stage-III dark 
energy experiment is about 5\% in $w_0$ and 30\% in $w_a$~\cite{soares}.  These surveys often have very stringent science quality requirements on their data products and also present a computing and data analysis challenge because of the large volume of data they produce.  The data products from these optical surveys,  such as photometric redshifts and cluster catalogs shall also be used  within Stage IV experiments including Euclid, WFIRST, etc.  To simplify the analysis of data from these 
surveys we have designed a flexible and automated data management system, which can be used to homogeneously process data from raw exposures to science data products and which can be used to compare the results from multiple surveys.

The outline of this paper is as follows. We shall briefly review one space-based stage IV dark energy experiment {\it eROSITA}, and highlight  its main science goals and discuss the current ground based surveys needed for science analysis. We then describe the key elements of our data management system and discuss how it is currently being used to process and calibrate survey data. Finally, we point out how this pipeline was used to process Pan-STARRS data to confirm galaxy cluster candidates from the Planck 2013 catalog.

\section{eROSITA on SRG}

The extended Roentgen Survey with an Imaging Telescope Array ({\it eROSITA})~\cite{predehl10,predehl14} is an X-ray telescope designed for an all-sky survey that will fly on the joint German-Russian Spectrum Roentgen-Gamma (SRG) mission to be launched in late 2016 from the  Baikonur cosmodrome into an L2 orbit. It consists of seven 1~deg$^2$ FOV telescopes with on-axis angular resolution of $\sim$15'' and energy sensitivity between 0.2 and 10 KeV. The main cosmology related science goals of {\it eROSITA} involve measurements with  galaxy clusters and large scale structure. {\it eROSITA} will discover more than 100,000 galaxy clusters out to redshifts beyond $z=1$ \cite{bib1}. The estimated accuracy in cosmological parameters from the {\it eROSITA} analysis is $\Delta \Omega_M = 0.0026$, $\Delta \sigma_8 = 0.0026$, $\Delta fNL=5.7$, and $\Delta w0=0.023$.  The corresponding dark energy figure of merit is about 300. More details on {\it eROSITA} technical specifications and science goals can be found in ref.~\cite{bib1}. To do cosmology with cluster number counts and the clustering of these clusters as a function of redshift, X-Ray data must be supplemented by optical data for each cluster to confirm it as a galaxy cluster and measure its photometric redshift.  In addition, we also expect to use optical data to do strong and weak lensing based mass measurements of {\it eROSITA} selected clusters. The list of optical photometric surveys,  which will be essential for followup of {\it eROSITA} in the sky region accessible to the German collaboration, along with some of their basic characteristics can be found in Table~\ref{tab:xxx}.

\begin{table}
\caption{An outline of some ongoing optical photometric surveys essential for optical followup of galaxy clusters detected by eROSITA in the western galactic hemisphere. These surveys were all designed with their own target science.}
\label{tab:xxx}
\smallskip
\centering
\begin{tabular}{|l|c|c|c|c|}
\hline
Survey & Start Date &  FOV(deg$^2$) & Coverage(deg$^2$) & Filters \\
\hline
PS1~\cite{PS1} & 2010 & 7  & 30,000  & $grizy$ \\
KiDS~\cite{KIDS} & 2011 & 1  & 1500 & $ugriz$ \\
ATLAS~\cite{ATLAS} & 2011 & 1 & 4500 & $ugriz$ \\
DES~\cite{DES} & 2013 & 3 & 5000  & $grizY$\\
HSC~\cite{HSC}  & 2013 & 1.5 & 1200 & $grizY$ \\
DECaLS~\cite{schlegel15} & 2014 & 3 & 6700 & $grz$ \\
\hline
\end{tabular}
\end{table}

\section{CosmoDM}

We now discuss the Cosmology Data Management System (CosmoDM) which has been used in processing and calibration of data from these optical photometric surveys.  The history of CosmoDM goes back to Dark Energy Survey Data Management system (DESDM), with development beginning in 2005, shortly after the submission of the successful DES proposal to NOAO.  DESDM was designed for the processing of data from DES.  DESDM has been extensively vetted through the start of DES survey using simulated DES data with annual data challenges\cite{Ngeow2006,Mohr2008,Mohr2012}. The development version of this pipeline has been used to process the data from the Blanco Cosmology Survey~\cite{BCS} as well as pointed optical follow-up data of South Pole Telescope Sunyaev-Zeldovich effect (SZE) galaxy cluster  candidates~\cite{Staniszewski,Song}. 

CosmoDM has been further developed at LMU in Munich since 2011 to prepare for external data calibration for Euclid and  {\it eROSITA}.  CosmoDM is designed to be flexible and robust enough to process data from a variety of telescopes with minimal changes to the processing software. The only change required to add a new camera comes to the software used in the first step of the single-epoch pipeline, where raw data are crosstalk corrected and altered to conform to the CosmoDM data model. However, CosmoDM can also use detrended data created using some other software as  an initial input, after converting them to our data model. CosmoDM consists of two main pipelines: (1) the single-epoch pipeline, or detrending pipeline, which processes and calibrates data from a single raw exposure and (2)  the co-addition pipeline, which combines single-epoch data (often from multiple nights) in the same part of the sky into a deeper image, which we call a ``coadd''. For processing surveys these coadds are usually constructed for $1 \times 1 $ square degree patches using a pre-defined grid on the sky, which we refer to  as  ``tiles'', but different tilings that, for example, focus on a specific set of targets with varying size can be easily configured. In the single-epoch pipeline we first apply corrections to remove various instrumental signatures. These include crosstalk, bias, flat-fielding, pixel scale and zeropoint flatness corrections.  Importantly, the output data products of the single-epoch pipeline have not been remapped to portions of tangent planes, and therefore the uncorrelated pixel noise is straightforward to track within an associated weight map.  Astrometric calibration is done using {\tt SCAMP}~\cite{scamp}, PSF modeling with {\tt PSFEx}~\cite{psfex} and model fitting photometry with {\tt SExtractor}~\cite{sextractor}.   We extract several hundred parameters for each detected object during the cataloging stage.  This finishes the single-epoch pipeline of CosmoDM.

Before we start the co-addition process, we carry out a relative photometric calibration of the single-epoch data for every coadd tile. 
To do this, we first assign an arbitrary zeropoint to one of the  images used to build the tile. We then solve for the ensemble of zero points 
for the remaining images which overlap the tile, by  using magnitude differences of stars from all overlapping pairs of images.  The selection of stars for this process is done using the new morphological star-classification parameter in {\tt SExtractor} called {\tt spread\_model}~\cite{BCS}.  Typically for each tile in surveys like DES, there are about 10 exposures (or $\simeq$ 300 images) per band for full-depth observations. For the DES data which consists of approximately 100 second exposures,  each CCD image  usually contains about 600  stars which are used to calculate magnitude differences.  At the end of this relative photometric calibration, every CCD image which overlaps the tile is assigned a constant  zeropoint.  We make no assumption about whether the night is photometric or not, and at present do not correct for spatial variation of the zeropoint within a CCD image due to integrated effects of clouds during the exposure.  We assess the quality of the relative zeropoint calibration using photometric  repeatability  scatter plots (of which an example is shown in Sect.~\ref{sec:ps1}). If the repeatability scatter is more than $\simeq$ 30~mmag, we proceed to identify and remove the poorly behaving image, and then repeat the calibration to bring the scatter below a target threshold.  This scatter threshold measured for bright objects provides an end to end measure of the systematic noise contributions in the system.  For DES data we find typical values below 10~mmag, and in the reanalysis of the PS1 data, we find typical scatter that is comparable.  More details of the relative photometric calibration procedure can be found in ref.~\cite{BCS}. 
 
Using these relative zeropoints,  we generate two kinds of coadds for a given region of the sky. The first type is called a non-homogenized coadd, where we directly combine the single-epoch images using the calibrated relative zeropoints with {\tt SWarp}~\cite{swarp}. These non-homogenized coadds are created mainly for cosmetic purposes for use in visualization of images and identification of clusters and brightest cluster galaxies (BCGs) within each cluster. The second type of coadd is PSF-homogenized.  In most of the current surveys, each object is observed $\simeq$~10 times in each band on different nights under different observing conditions. This generically leads to sharp position variations in PSF in the coadd in regions corresponding to the boundaries of the underlying single-epoch images.  For example, the eastern and western portion of a single star can exhibit very different FWHM.  These PSF discontinuities affect star-galaxy separation, contribute to position dependent completeness variations and make it generally impossible to apply PSF-corrected model fitting photometry to the coadd image 
(See however ~\cite{Tyson}, who perform a  PCA based reconstruction of the PSF for individual CCD images and then make use of these PSFs during co-addition. They demonstrate an excellent ability to track the residual PSF ellipticity and show that the galaxy ellipticity correlation is consistent with shot noise.)
To alleviate these issues, we transform the  single-epoch images within a single band and on a single coadd tile (typically 62$'\times$62$'$) to a common, Moffat PSF using the code {\it PSFNormalize}~\cite{BCS} with input position dependent PSF transformation kernels from {\it PSFEx}.  We homogenize to the median PSF of all the single-epoch images contributing to a band.  These PSF normalized single-epoch images, which have the same PSF, are then co-added using {\tt SWarp} with flux scales obtained from the relative internal photometric calibration.  These coadd images with a common PSF at all locations are termed PSF-homogenized coadds.  A disadvantage of PSF homogenization however is that there is a degradation of the best seeing images and Poisson noise is amplified for the worst seeing images. The advantage is that PSF corrected model fitting photometry is possible on these image.  More details of the PSF homogenization procedure are provided in ref.~\cite{BCS}.  

\begin{figure}[tbp] 
\centering
\includegraphics[width=.6\textwidth]{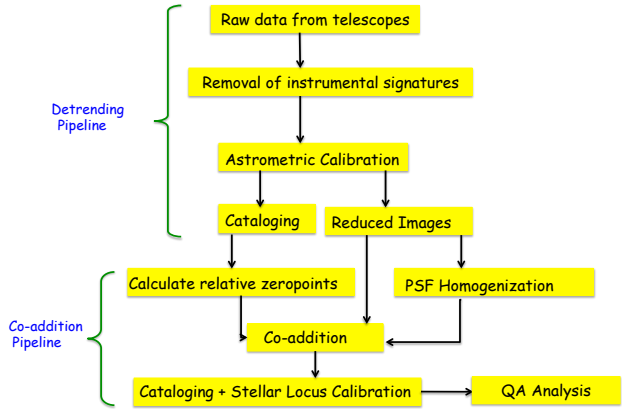}
\caption{Flowchart of the single-epoch and co-addition processing and calibration within CosmoDM}
\label{fig:cosmodmflow}
\end{figure}

The PSF homogenized coadds are then cataloged with PSF-correcting, model-fitting photometry using {\tt SExtractor} in dual image mode where typically the detection image is a $\chi^2$ image \cite{szalay99} composed of $r$+$i$ or $i$+$z$.  The absolute photometric calibration at the end of the co-addition pipeline is done using the position of the stellar locus~\cite{SLR} together with the 2MASS J-band to get the absolute zeropoint~\cite{BCS,Song,Liu}.  The flowchart of the single-epoch and co-addition pipeline in CosmoDM is shown in figure~\ref{fig:cosmodmflow}. 

CosmoDM processing is done on supercomputers, where we take advantage of the large number of cores, by distributing the jobs to various multi-core nodes and using the parallel file system. Astronomical data processing is usually thought of as high-throughput computing because of the large data volumes and because the disk I/O is a limitation to the scaling on a high performance computing platform.  Since CosmoDM contains many standalone executables and because different stages of the processing and calibration require different combinations of data, it is non-trivial (and probably impossible) to create one library or executable which would hold all image data in memory from the beginning to the end. Therefore, to enable scaling to a larger number of parallel jobs within a particular computing environment, we store intermediate data products on RAMdisk and we read and store only RICE tile compressed FITS images~\cite{Pence}.  

Although detailed benchmarking studies of the benefits of RAMdisk are still ongoing, from our preliminary tests  we find that reading from RAMdisk is about 100 times faster than from the parallel filesystem consisting of underlying RAID arrays of hard disks. While doing lossy (for floating point images) Rice Compression, we  use the CFITSIO default quantization factor of four. This means that the spacing between quantized integer levels is equal to the RMS noise divided by a factor of four. In addition to this, dithering is also applied, which is a process that adds noise to minimize systematic biases. The net increase in noise with this compression algorithm is only 0.26\%~\cite{Pence}.  With these settings we obtain a typical reduction in file size of about a factor of five. More details of tile compression used in CFITSIO and the loss in precision as a function of compression factor can  be found in refs.~\cite{Pence,Pence1} and references therein. According to our preliminary estimates, we find that although opening a tile-compressed image is faster by a factor of two, loading the full image data in memory takes about twice the amount compared to a regular fits images. Thus, there is no net I/O time benefit in storing intermediate data products as tile compressed images.  However, because of the factor of five compression in the FITS file,  we can store more images in RAMdisk during the pipeline processing and the disk I/O is reduced also by a similar factor.  Our final data products are also stored as tile-compressed images.  Together, the RAMdisk based pipeline and the RICE tile compression allow us to reduce the overall disk I/O load by a factor of $\sim$50.  For a given supercomputer with a fixed parallel filesystem bandwidth, this translated into an ability to run $\sim$50 times more processing jobs simultaneously.  This improvement factor translates directly into reduced reprocessing times for large surveys.

On our typical platform at the Computational Center for Particle and Astrophysics (C2PAP), our system uses a single 32 core node to process a single input DECam (3~deg$^2$) exposure in $\sim$10 minutes, and we have sustained runs using approximately 60 nodes over two weeks to process, for example, the year 1 DES survey data (corresponding to $\sim$50,000~deg$^2$ of science imaging in all bands combined).

The job submission in CosmoDM is done using Condor-G~\cite{condor} and the WS-gram~\cite{globus} interface, which allows us to run on supercomputers with diverse batch job submission queues. Currently we have deployed and run CosmoDM on three different computing clusters with Slurm, LoadLeveler and PBS batch job submission queues.   We use an ORACLE database to  store all our image and processing metadata  as well as the object catalogs.  Data transfer is accomplished using the Globus tookit~\cite{globus}, as it is faster than ordinary ftp or scp, and a structured hierarchical archive where each data product has a combination of associated identification tags that uniquely identify its position within the archive.  

A novel feature of CosmoDM is that we made the widely used Astromatic codes such as {\tt SExtractor}, {\tt SCAMP}, and {\tt SWarp} CFITSIO compatible, which allows us to directly read/write tile-compressed images.  We are developing a Quality Assurance (QA) based framework to create detailed data quality information during every step of the processing to check that a given data product satisfies all data quality tests. We are also gearing up for additional scaling tests on the SuperMUC supercomputer to understand the maximum number of single-epoch exposures we can process and calibrate in parallel without significant impact on the parallel file system after switching to RAMdisk to store intermediate data products. The science codes and QA framework from CosmoDM is useful now as a prototype for the ground-based external dataset calibration pipeline needed for  Euclid.

As of Jan. 2015, we have successfully processed data from four different surveys/telescopes with CosmoDM.  We briefly list the different datasets analyzed including the science goals. 
\begin{itemize}
\item Dark Energy Survey data.
\item DECam data for optical followup of XMM-XXL field
The XMM-XXL survey~\cite{XXL} is a XMM based survey which has mapped out  two extragalactic regions of 25 deg$^2$ using 6.9 megaseconds of observing time. This field was observed with DECam in July-August 2013 in $griz$ bands. The data was processed with CosmoDM, and joint analysis of this dataset together with the X-ray data is
in progress. 
\item CFHTLS MegaCam data. We processed  the CFHTLS Megacam data in $griz$ bands starting with the detrended T0006~\cite{CFHT} data in the COSMOS field and created coadds.  We identified problems with the T0006 dataset and communicated these to the project.   The final goal is a comparison with the TeraPix reductions. We are now awaiting the release of the improved T0007 single-epoch data.
\item Pan-STARRS data. This is described in the next section.
\end{itemize}

\section{Processing of Pan-STARRS data}
\label{sec:ps1}
As a warmup to our Pan-STARRS follow-up program of {\it eROSITA} sources, we undertook a pilot project  with SZE selected cluster candidates from the Planck SZE survey.  We briefly describe this project here, but more details are provided in ref.~\cite{Liu}.  In March 2013, the Planck Collaboration released an SZE source catalog consisting of about 1200 galaxy cluster candidates~\cite{Planck}.  To confirm these as galaxy clusters and measure their photometric redshifts, optical data are needed.  As members of the Pan-STARRS collaboration, we decided to examine those as-yet unconfirmed candidates that overlap the 3$\pi$ survey of the Pan-STARRS1 (PS1) survey~\cite{PS1}.  

The PS1 3$\pi$ survey is carried out with 1.4~Gpixel CCD camera with a 7 deg$^2$~FOV on a 1.8~m telescope on Haleakala in $grizy$ filters.  The focus of the 3$\pi$ survey is to image the entire sky with $\delta > -30^{\circ}$ to depths as deep as or deeper than SDSS~\cite{sdss}. The PS1 filter set and depth is sufficient to measure cluster redshifts over a wide redshift range, extending out to $z\sim0.6$. 

Our procedure for the preparation of the calibrated science catalogs using CosmoDM is slightly different than outlined in Fig.~\ref{fig:cosmodmflow}. Instead of starting from raw data,   the initial input data products  are remapped single-epoch exposures (called ``warps'')  from the PS1 collaboration in Hawaii which were created using Version 2 (PV2) of the IPP pipeline~\cite{Metcalfe}. We  use CosmoDM to create coadds and associated PSF-corrected model fitting catalogs for about 400 cluster candidates.  The first step is to obtain PSF-corrected model fitting catalogs for all the input warps and to use these catalogs in the overlapping images to determine the relative zeropoints of these images.  These warps are then processed using the co-addition pipeline into calibrated science ready coadd images and catalogs in the same manner as described earlier.

\begin{figure}[tpb]
\centering
\includegraphics[width=.5\textwidth]{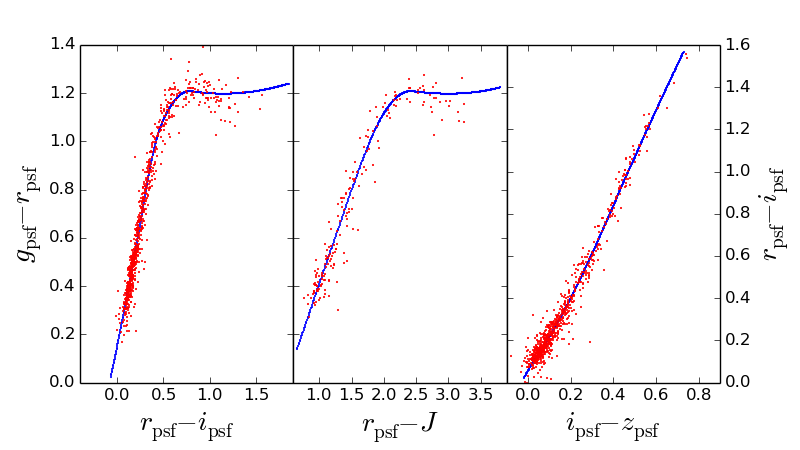}
\caption{Stellar Locus plot for Planck cluster candidate $507$. The blue line shows the PS1 stellar locus and red points show the PSF magnitudes for stars.
The scatter around the stellar locus for this tile is 27 mmag, 19~mmag and 53 mmag in  $g-r$ vs $r-i$ (left panel), $r-i$ vs $i-z$ (right panel) and 
$g-r$ vs $r-J$ (middle panel) respectively.}
\label{slrplot}
\end{figure}

\begin{figure}[tpb] 
\centering
\includegraphics[width=.5\textwidth]{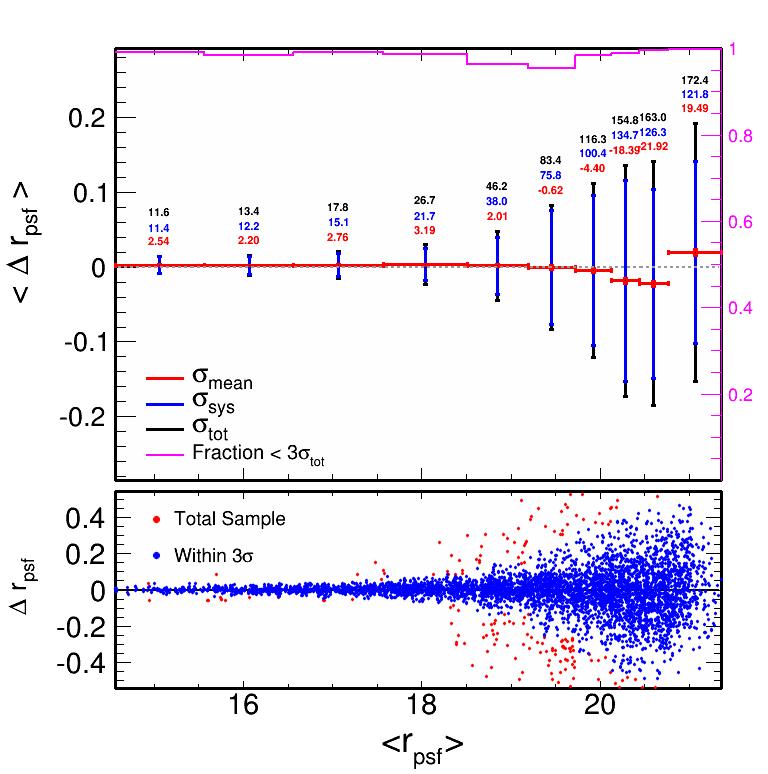}
\caption{Repeatability scatter plot for Planck cluster candidate $706$ in $r$ band using $mag\_psf$. The top panel shows the RMS magnitude difference between single-epoch images which spatially overlap with each other, as a function of magnitude along with the systematic (blue) and total (black) RMS scatter. The lower panel shows a scatter plot with individual stellar pairs.  The total repeatability scatter at the bright end is about 11 mmag. The full distribution of the scatter for all Planck cluster candidates in $griz$ bands is shown in Ref.~\cite{Liu}.}
\label{fig:rep}
\end{figure}

\begin{figure}[tpb] 
\centering
\includegraphics[width=.5\textwidth]{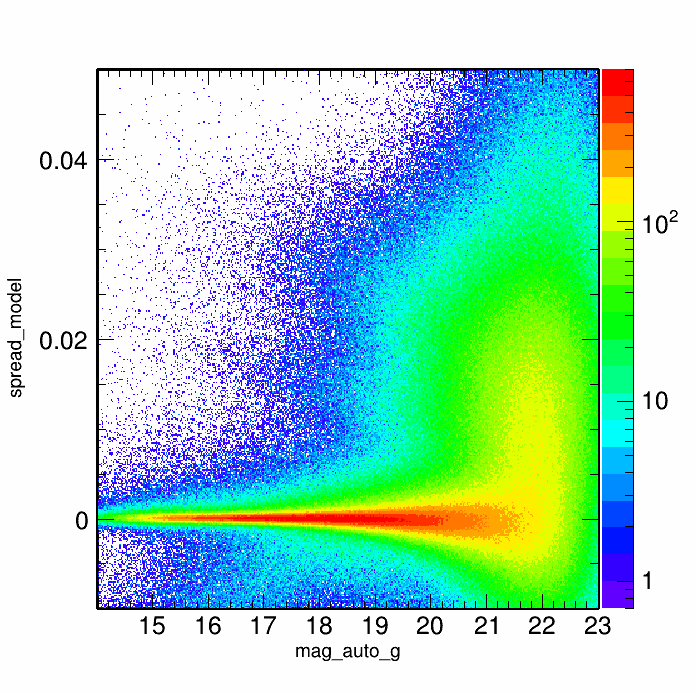}
\caption{Distribution of a morphological based star-galaxy separator in {\tt SExtractor} called {\tt spread\_model}~\cite{BCS} for PS1 data around Planck cluster candidates as a function of magnitudes. One can see the narrow stellar sequence near values of 0.0 and the galaxy cloud extending to more positive values. Objects with {\tt spread\_model} $< -0.002$ are junk objects as discussed in ref.~\cite{Bertin13}}
\label{fig:xxx}
\end{figure}

 The data quality indicators for our PS1 coadd images are quite good.  We obtain a stellar locus scatter of 33~mmag, 24~mmag, and 58~mmag in $gr-ri$, $ri-iz$, and $gr-rJ$ color spaces, respectively, which is somewhat better than the scatter in the same color spaces that we obtain when using SDSS data as input~\cite{BCS}. The stellar locus scatter plots for different color combinations are shown for one Planck cluster candidate $507$ in Fig.~\ref{slrplot}.  The repeatability scatter, which measures how similar the photometric measurements of the same star are during repeat visits in the survey is about 16-18 mmag in $griz$.  We show the repeatability plot for one such Planck cluster candidate $706$ in $r$ band in Fig~\ref{fig:rep}. The repeatability scatter for this coadd tile in $r$ band is about 11~mmag. The corresponding internal photometric accuracy reported from the Pan-Starrs Ubercal analysis of data taken during photometric periods is about 12~mmag in $griz$~\cite{Schlafly}. 
  
 The confirmation of galaxy cluster candidates is done using catalogs to probe for peaks in the galaxy density along the line of sight toward each candidate as a function of the expected color of the red sequence at each redshift.  Candidates are confirmed by visual inspection using pseudo-color images constructed from the non-homogenized coadds.  We confirm 60 new galaxy clusters and measure photometric redshifts with a characteristic accuracy of $\sigma_{\frac{\delta z}{1+z}} \simeq 0.022$~\cite{Liu}. For this analysis, we  used the {\tt SExtractor} measurements of  $mag\_auto$ for total magnitudes and the newly available  $mag\_detmodel$  for galaxy colors.

One element of the system that enables both good photometric calibration and also good selection of galaxies for cluster studies is the new morphological classifier {\tt spread\_model}, which has been added to {\tt SExtractor} and is extracted now within the co-addition pipeline.   Figure~\ref{fig:xxx} contains the distribution of {\tt spread\_model} as a function of $g$ band magnitude for all PSF-homogenized coadds used in the study.  One can see the  stellar sequence in $spread\_model$ at values near 0, and the galaxy cloud extends to higher positive values.  The stellar sequence and the galaxy cloud are quite distinct until magnitudes of $g\sim20$, where they begin to overlap significantly due to noise in these 3$\pi$ survey coadds.  In general, when using {\tt spread\_model} with PSF homogenized coadds, the performance is far better and enables morphological classification to much fainter magnitudes than the earlier {\tt class\_star} parameter.  More details are available in ref.~\cite{BCS,Bertin13}.

We are now using CosmoDM and applying this same approach to the processing of about 8000~deg$^2$ of PS1 that overlaps the western galactic hemisphere in preparation for the upcoming {\it eROSITA} launch as well as for additional studies of cluster samples extracted from existing X-ray data. 

\section{Conclusions}

We have designed an automated data management system (CosmoDM) which can process in a homogeneous way,  data from modern astronomical imagers starting from raw data up to the creation of  science ready data products.  CosmoDM will be used to process optical data from ongoing and recently completed surveys as Dark Energy Survey, Pan-Starrs, etc, which are needed to complement X-ray measurements of galaxy clusters from {\it eROSITA}.  As of now, we have deployed  CosmoDM on multiple supercomputers and successfully processed  data from  DES, CFHT, and Pan-Starrs. We have worked to optimizing the I/O load and the processing time as well as to improve the data quality of the processed data. As a pilot project for our {\it eROSITA} followups, we have used the current development version of CosmoDM to process detrended and astrometrically calibrated Pan-Starrs data for about 400 galaxy cluster candidates detected by Planck. Using the stacks, we have been able to confirm 60 of these clusters and measure red-sequence cluster redshifts with accuracy of about 2\%.

\acknowledgments
We would like to thank the anonymous referee for detailed comments and suggestions which helped us improve this paper. We are grateful to Bill Pence and Rob Seaman for many helpful discussions and fixes  related to FITS  tile compression for the past seven years. We acknowledge the support by the DFG Cluster of Excellence ``Origin and Structure of the Universe'' and the Transregio program TR33 ``The Dark Universe''.  We thank the DLR for its support of the LMU Euclid program. The data processing has been carried out on the computing facilities of the Computational Center for Particle and Astrophysics (C2PAP).

\end{document}